\documentclass[twocolumn,aps,preprintnumbers,prl,showpacs,amsfonts,amsmath,amssymb,superscriptaddress,floatfix]{revtex4}
\usepackage[]{verbatim}
\usepackage{graphicx}
\usepackage{color}
\usepackage{braket}
\allowdisplaybreaks[3]

\begin{document}
\title{Coupling single emitters to quantum plasmonic circuits}
\date{\today}
\author{Alexander Huck}\email{alexander.huck@fysik.dtu.dk}
\affiliation{Department of Physics, Technical University of Denmark, Fysikvej, Building 309, 2800 Kgs. Lyngby, Denmark}
\author{Ulrik L. Andersen}\email{ulrik.andersen@fysik.dtu.dk}
\affiliation{Department of Physics, Technical University of Denmark, Fysikvej, Building 309, 2800 Kgs. Lyngby, Denmark}

\begin{abstract}
In recent years the controlled coupling of single photon emitters to propagating surface plasmons has been intensely
studied, which is fueled by the prospect of a giant photonic non-linearity on a nano-scaled platform. In this article
we will review the recent progress on coupling single emitters to nano-wires towards the construction of a new platform
for strong light-matter interaction. The control over such a platform might open new doors for quantum information
processing and quantum sensing at the nanoscale, and for the study of fundamental physics in the ultra-strong coupling
regime.
\end{abstract}

\pacs{} \maketitle

\section{Introduction}
When directing a photon towards a single emitter, e.g. an atom, a quantum dot (QD) or a defect center, the probability
that the photon being absorbed (or emitted in a single mode) is typically very low due to the small cross section of
the emitter~\cite{2000Loudon}. By strongly focussing the photon, the absorption probability can be largely
improved~\cite{2001Enk,2007Wrigge,2008Tey} (in particular if the photon is radially polarized~\cite{2007Sondermann}),
but reaching high probabilities is highly challenging in particular if the emitter is inhomogeneously
broadened~\cite{1964McCumber}. Another common approach to enhance the photon-emitter interaction strength is to place
the emitter inside a cavity in which the photon bounces back and forth several times thereby increasing the probability
for absorption~\cite{2003Vahala}. This approach, which is related to the Purcell effect~\cite{1946Purcell}, however can
be only used for narrow band emitters and photons due to the small bandwidth of the cavity. Finally, strong absorption
can be attained by placing the emitter at an optimal position in the confined field of the plasmonic mode propagating
along a small waveguide made of metal. As a plasmon mode, the propagating eigenmode of a waveguide made of metal, can
be tightly confined to below the optical diffraction limit~\cite{1997Takahara}, the waveguide acts as an ultra-strong
lens that focusses the light down to a few nanometers in the transverse dimension. Due to this exceptionally strong
focussing capability, the emitter can interact with the photon with an extra-ordinary strength and thus absorb the
photon with unit probability, or equivalently, a photon emitted from the dipole can be directed into a single plasmonic
mode of the wire with unit probability~\cite{2006Chang}.

\begin{figure}[hbt]
    \begin{center}
    \includegraphics[width=0.8\columnwidth]{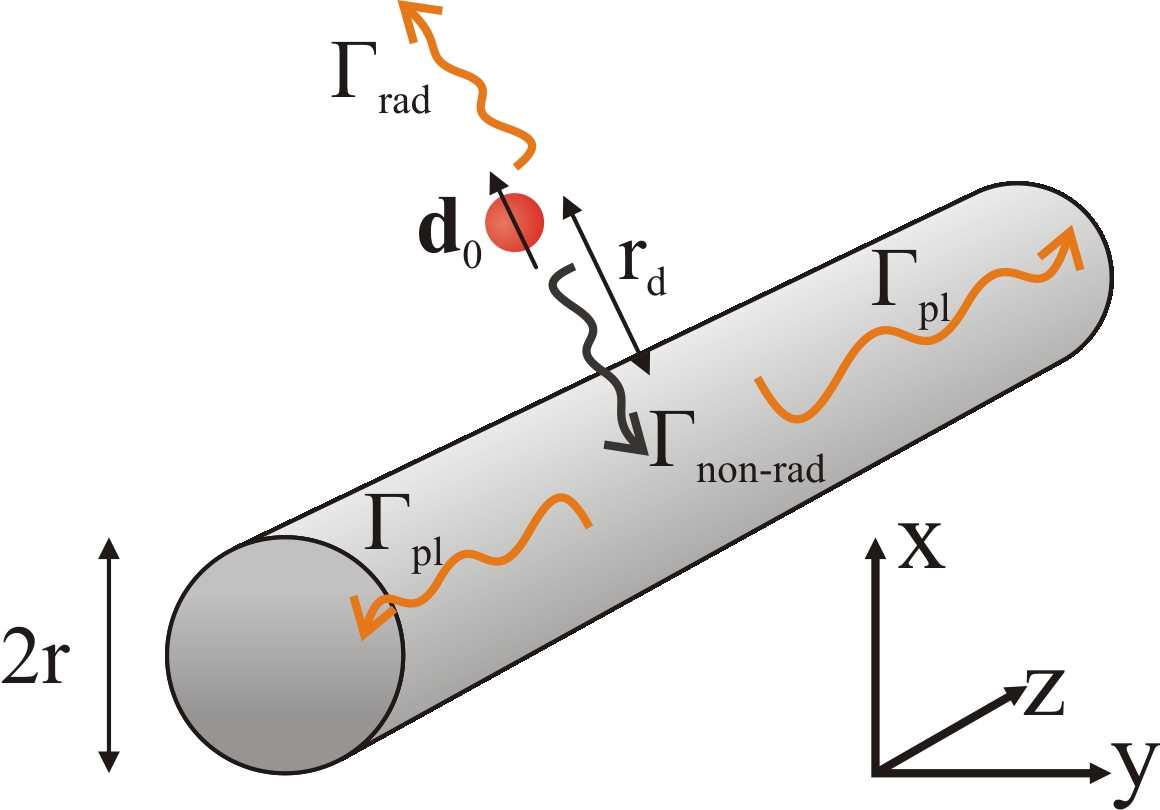}
    \end{center}
    \caption[]{Schematic illustration of the individual decay channels of a dipole emitter $\mathbf{d}_0$ located
    at a distance $\mathbf{r}_d$ next to a metallic nanowire. The emitter (orange sphere with black arrow)
    decays by emitting a single photon which is either directed into free space, into ohmic losses of the metallic
    wire, or into the single plasmonic mode propagating along wire. $\Gamma_{rad}$, $\Gamma_{pl}$, and
    $\Gamma_{non-rad}$ denote the decay rates to the radiation field,
    the propagating surface plasmon mode, and the metallic loss channels, respectively. The ratio between the different
    rates is determined by the distance between the dipole emitter and the wire, $\mathbf{r}_d$. \label{fig-illustration}}
\end{figure}
Surface plasmon polaritons are electromagnetic excitations of charge density waves at the interface between a conductor
and a dielectric medium and, as mentioned above, they can be confined to transverse dimensions much smaller than what
is possible with conventional optics~\cite{2007Maier,2010Gramotnev}. Applying Fermi's golden rule to evaluate the
interaction strength between a dipole transition and such a surface plasmon mode, the coupling strength $g$ is found to
scale with the local electric field per photon, $\mathbf{E}_0(\mathbf{r}_0),$ which is higher when the field
confinement increases~\cite{2006Chang}. About a decade ago, it was theoretically suggested that the resulting coupling
rate to propagating surface plasmon modes, $\Gamma_{pl},$ can largely exceed the spontaneous emission rate to the
radiation field, $\Gamma_{rad},$ and that non-radiative decay rates, $\Gamma_{non-rad},$ are negligible for the
considered parameter space~\cite{2006Chang} \footnote{Note that although this channel is coined non-radiative, it is in
fact radiative but the radiation is not accessible as it goes into ohmic channels of the wire. It is not connected to
non-radiative transitions of the emitter.}. A schematic illustration of the system, as an example we chose a nanowire
plasmon waveguide, and the rates are illustrated in Fig.~\ref{fig-illustration}. These findings stimulated an enormous
interest and triggered experimental efforts in utilizing surface plasmon modes for enhanced light matter interaction
and device applications. It is worth mentioning that propagating surface plasmons are different from localized surface
plasmons. In contrast to localized plasmons, propagating plasmons are traveling in a well-defined direction as the real
part of their wavevectors are nonzero. This basically means that the propagating plasmon can be guided through metallic
circuits much like optical beams in photonic circuits. The difference between the photonic and plasmonic guiding,
however, is that the transverse dimension of the latter guiding can go below the diffraction limit and thus interact
strongly with single emitters. The vision is therefore to build arbitrarily complex plasmonic circuits with metallic
waveguides in which the plasmons can strongly interact with emitters, and thereby enable classical and quantum
information processing with a very small foot-print. In addition to information processing, a scalable plasmonic
platform of strongly interacting emitters and photons might be a promising system for investigating out-of-equilibrium
many-body physics.
\\
In this Review, we will describe experimental efforts towards the realization of an efficient interaction between a
single optical dipole emitter and propagating surface plasmon modes supported by metallic waveguides. There exists a
vast literature on plasmonics systems and circuits which have been discussed in a number of review
articles~\cite{2010Gramotnev,2015Fang,2012Sorger,2011Benson,2012LeonRev,2013Tame}. In the present Review article,
however, we mainly focus on quantum plasmonic circuits in which quantum states, and in particular single plasmonic
states, are propagating. We thus focus on reviewing work on which single propagating plasmons are generated by a single
emitter. We start by discussing the definition of \textit{quantum plasmonics} and subsequently summarize different
potential applications of the plasmon-emitter platform. Then we discuss the various experimental activities solely
focussing on the coupling between a \textit{single emitter} and a propagating plasmon supported by a metallic
waveguide. We conclude with an outlook in which we summarize a number of challenges that have to be solved for further
progress and new directions of the field.

\section{Quantum plasmonics}
The property of strong confinement of the electro-magnetic field has opened a new world of opportunities both in
classical and quantum optics. Applications in the latter field have triggered the new field of quantum
plasmonics~\cite{2013Tame}. But what is "quantum" in quantum plasmonics? Before answering this question, let us briefly
review some of the first experiments carried out at the interface between quantum optics and plasmonics.

The experiment reported by Altewischer et al.~\cite{2002Altewischer} originally triggered curiosity of the quantum
optics community in the field of plasmonics. They have shown that the entanglement present in the polarization degree
of freedom of two spatially separated optical modes survived after one of the photons passed through a gold film
perforated with nano-meter size holes. In the holes, photons excite surface plasmon resonances and thus enhance the
total transmission through the gold film~\cite{2002Ebbesen}. Later experiments reported by Fasel et
al.~\cite{2005Fasel} confirmed these findings by demonstrating that also time-energy entanglement was preserved after a
photon - surface plasmon - photon conversion process. The survival of quadrature squeezing after surface plasmon
excitation demonstrated that the phase coherence is also not affected by surface plasmon excitations~\cite{2009Huck,
2013Lawrie}, and that linear propagation losses can be modeled by an effective beam splitter interaction. Together with
the demonstration of Hong-Ou Mandel interference of identical photons (produced externally by spontaneous parametric
down-conversion) on a surface plasmon based beam splitter~\cite{2014Fakonas, 2014Martino, 2013Heeres}, this series of
experiments leaves little doubt that surface plasmon modes are well described by bosonic field operators.

These experiments suggest that plasmons behave similarly to bosonic electro-magnetic fields. This is also what is
expected from theory derived by Bohm and Pines in the 1950s~\cite{1953Pines}, establishing the quantization of
collective electron oscillations which have bosonic properties. Due to the large number of electrons, a macroscopic
collective excitation is formed and can be described by the macroscopic permittivity $\epsilon$ of the materials. This
permittivity determines the mode profile and their guidance along a waveguide~\cite{1997Takahara}, and in a modern
context the quantization of surface plasmon modes has been addressed by Tame et al.~\cite{2008Tame} taking metal losses
into account.

The afore mentioned experiments are referred to as being \textit{quantum plasmonics} experiments due to the fact that
the statistics of the propagating plasmons cannot be described by a well behaving Glauber-Sudharshan P-function. This
function is defined as a weight function, $P(\alpha),$ in the coherent state basis $\ket{\alpha},$ where $\alpha$ is
the complex coherent state amplitude. The density matrix $\hat{\rho}$ of a single mode state can then be written as
$\hat{\rho} = \int P(\alpha) \ket{\alpha}\bra{\alpha} d^2\alpha$~\cite{1963Glauber, 1963Sadurshan}. Quadrature squeezed
states, single photon states and other non-Gaussian pure states do not possess a well behaving P-function and are thus
often considered as non-classical states. Therefore, the excitation of plasmonic modes in such non-classical bosonic
modes are often coined \textit{quantum plasmonics}~\cite{2013Tame}.

However, there exists also another definition of \textit{quantum plasmonics} which is relevant when the structures are
reduced to very small dimensions of around a few nanometers. In this regime, the standard assumption of a continuous
energy spectrum of the electrons might break down as the electrons become bounded and thus exhibit a quantized energy
spectrum~\cite{1986Halperin}. Furthermore, if very short length scales are considered, the quantum delocalization
nature of electrons might be relevant and should be taken into account~\cite{2015Esteban}. Due to this possible need
for taking into account the energy quantization and the non-local nature of the electrons, the plasmonic behavior can
only be described by a full quantum model and thus it is refereed to as \textit{quantum plasmonics}~\cite{2013Tame}.
However, all experiments to date on propagating plasmons have not been affected by these quantum phenomena (to the best
of our knowledge) and thus the "quantum" in \textit{quantum plasmonics} in previous experiments simply refers to the
quantum statistics of the propagating bosonic mode. In addition to the entangled and squeezed plasmonic states
discussed above, there has been a number of activities devoted to the generation of single plasmonic states by coupling
single emitters to metallic waveguides. This will now be discussed in greater details.

\section{Emitter coupling to surface plasmons}
In this section, we first consider the coupling of an emitter to a plasmonic mode supported by a metallic waveguide, as
illustrated in Fig.\ref{fig-illustration}. The emitter with an angular transition frequency $\omega$ and dipole moment
$\mathbf{d}_0$ is positioned a distance $\mathbf{r}_d$ away from the surface of the waveguide. The one-dimensional
waveguide is considered to have finite dimensions in the transverse (x,y)-plane and to be infinite in the longitudinal
z-direction. The original framework for emitter - waveguide mode coupling was set by Klimov et al.~\cite{2004Klimov}.
Using the quasistatic approximation they derived an analytical expressions for the spontaneous emission rate into the
guided mode, $\Gamma_{pl}$,the radiation field, $\Gamma_{rad},$ and non-radiative emission due to dissipation of the
fiber, $\Gamma_{non-rad}$, which is related to the imaginary part of the permittivity, $\mathfrak{Im}(\epsilon)$. This
work was later extended by Chang et al.~\cite{2006Chang} to the case of metallic nanowires supporting propagating
surface plasmon modes~\cite{1997Takahara}. It was found that for an optimal dipole orientation parallel to the electric
field component of the radially polarized surface plasmon mode, $\mathbf{d}_0
\parallel \mathbf{E}_r ,$ the decay rate into the plasmon mode $\Gamma_{pl}$ (the guided mode in the dielectric
nanofiber case) can largely exceed the sum of all other decay channels, i.e. $\Gamma_{pl} \gg \Gamma_{rad} +
\Gamma_{non-rad}$. This results in a $\beta-\mathrm{factor}$ very close to 1, where the $\beta-\mathrm{factor}$ is
defined as the ratio of emission going into the guided modes to the total decay rate, $\beta = \Gamma_{pl} /
\left(\Gamma_{pl} + \Gamma_{rad} + \Gamma_{non-rad}\right)$. This result is the central motivation for the experimental
investigations discussed in the following sections. The strong $\Gamma_{pl}$ component, relative to all other decay
channels, effectively originates from the tight mode confinement associated with the fundamental surface plasmon mode.
For small $\mathbf{r}_d$ the $\Gamma_{non-rad}$ component dominates since it scales with $1/\mathbf{r}_d^3$, reflecting
the near field of the emitter~\cite{2006Chang}.

If the plasmon mode can not be described analytically, for example when the nanowire is placed on a substrate or in
case of other waveguides such as grooves or wedges, the total decay rate $\Gamma_{tot}$ and the individual decay
channels ($\Gamma_{pl},$ $\Gamma_{non-rad},$ and $\Gamma_{rad}$) can only be obtained numerically using a
finite-element method~\cite{2010Chen} or a finite difference time domain method~\cite{2007Kaminski}. While for the
normalized plasmonic decay channel $\Gamma_{pl}/\Gamma_{0}$ it is sufficient to know the electric $\mathbf{E}(x,y)$ and
magnetic $\mathbf{H}(x,y)$ field distributions in the transverse $(x,y)$-plane, a rigorous modeling in three-dimensions
is required to obtain the total decay rate $\Gamma_{tot}$ and to estimate the $\beta - \textrm{factor}.$

\begin{figure}[hbt]
    \begin{center}
    \textbf{Barthes et al., 2011}
    \end{center}
    \caption[]{Plot of the modified two-dimensional density of states, $\Delta \rho_\mathbf{u}^{2D}$, as a function of
    normalized longitudinal wavenumber $k_z/k_0,$ where $k_0$ is the wavenumber of the surrounding, for a dipole
    located a distance $d$ from a cylindrically shaped silver nanowire with a radius of $20 \, nm$~\cite{2011Barthes}.
    The dipole is radially oriented with respect to the wire surface and located in a homogeneous material
    with $\epsilon = 2$. The contributions of $\Delta \rho$ to
    $\gamma_{rad}, \, \gamma_{pl}$ and the non-radiative modes $\gamma_{NR}$ are indicated at the bottom of the figure. The
    large contribution of the guided plasmon mode is highlighted by the peak centered around the effective mode index
    $n_{eff}=2.28$ of the plasmon mode.
    \label{fig-Greens-dyad}}
\end{figure}

One may also obtain the decay rate of a dipole emitter at a position $\mathbf{r}$ using the proportionality
$\gamma(\mathbf{r}) \propto \mathfrak{Im} Tr \mathbf{G}(\mathbf{r},\mathbf{r})$, where
$\mathbf{G}(\mathbf{r},\mathbf{r})$ is the three-dimensional Green's tensor~\cite{2006NovotnyHecht}. In the vicinity to
a two-dimensional, infinitely long waveguide with arbitrary shape, $\mathbf{G}(\mathbf{r},\mathbf{r})$ can be expressed
in a Fourier serious of two-dimensional Green's tensors
$\mathbf{G}^{2D}(\mathbf{r}_{\parallel},\mathbf{r}_{\parallel},k_z),$ where $k_z$ is the longitudinal wavenumber, as
done by Barthes et al.~\cite{2011Barthes}. The two-dimensional Green's tensor is then further separated as
$\mathbf{G}^{2D} = \mathbf{G}^{2D}_{ref} + \Delta \mathbf{G}^{2D},$ where $\mathbf{G}^{2D}_{ref}$ accounts for the
contribution from the reference system such as the homogeneous background or the substrate and $\Delta \mathbf{G}^{2D}$
describes the contribution from the waveguide structure. The waveguide contribution to the modification of the decay
rate, expressed as a function of $k_z$, is then obtained as $\Delta \rho_\mathbf{u}^{2D} (\mathbf{r}_{\parallel},k_z)
\propto \mathfrak{Im} \left\{ \epsilon(\mathbf{r}_{\parallel}) Tr\left[ \mathbf{u} \cdot \Delta \mathbf{G}^{2D} \cdot
\mathbf{u} \right] \right\},$ where $\mathbf{u}$ is a unit vector along the direction of the dipole. $\Delta
\rho_\mathbf{u}^{2D}$ is referred to as the modified two-dimensional local density of states and plotted in
Fig.~\ref{fig-Greens-dyad} as a function of $k_z$ for a cylindrically shaped nanowire made of silver and a radially
oriented dipole. Partial integrations of $\Delta \rho_\mathbf{u}^{2D}$ then yield the contributions from the total
emission rate to the radiation field, the guided plasmonic mode, and lossy plasmonic modes (the non-radiative channel),
as indicated in Fig.~\ref{fig-Greens-dyad}.

\section{Experimental approaches}
\subsection{Requirements on the optical setup}
Optical investigations on surface plasmon emitter coupling are conveniently carried out with a confocal microscope. A
high optical resolution better than $1\, \mu m$ and a large photon collection efficiency are general requirements on
the setup. Both the resolution and the collection efficiency scale with the numerical aperture (NA) of the microscope
objective. An excitation laser should be chosen to match the absorption band of the emitter. Avalanche photodiodes
(APDs) with low dark noise (usually $< 1000 counts/s$) are essential to acquire signals from single emitters. For an
evaluation of the total coupling strength the emitter lifetime $\tau$ has to be determined requiring a pulsed laser
with a pulse width much smaller than the emitter lifetime. The minimum detectable lifetime $\tau_{min}$ is roughly
given by the instrument response function of the setup, which is usually limited by the APD time jitter of $50-300 \,
\text{ps}$ depending on the model.

\subsection{Metallic waveguides supporting plasmon modes}
The choice of metal and the shape of the waveguide are central design aspects when considering emitter surface plasmon
coupling. Due to the small $\mathfrak{Im}(\epsilon_{Ag}) $ ($ \approx 1 - 3$ for wavelengths between $600 \, nm$ and $1
\, \mu m$) silver is the preferable material for emitters operating in the visible and near-infrared spectral range.
However, unprotected silver corrodes in ambient air which makes it important to pay special attention on the
fabrication method. For wavelengths in the near-infrared part of the spectrum also gold ($\mathfrak{Im}(\epsilon_{Au})
\approx 3.5 - 10 $) may be used. Gold is chemically stable and therefor ensures long time operation of the structure.

It was encountered that nanowires made of silver or gold fabricated with electron-beam lithography and thermal or
electron-beam assisted deposition of metals bear several drawbacks for emitter plasmon coupling, mainly because the
resulting metallic structures are poly-crystalline, i.e. they are composed of particle clusters with individual
particle sizes in the nanometer range. Hence, lithographically prepared nanowires show increased surface plasmon
propagation losses due to scattering associated with the inherent roughness of the structure, compared to colloidal
nanowires prepared with a wet-chemical method~\cite{2005Ditlbacher}. Clusters of silver nano-particles may fluoresce
when illuminated with laser light~\cite{2001Peyser} and in gold, electrons can be excited from the d-band above the
Fermi level and afterwards recombine radiatively with a small efficiency ($\sim 10^{-10}$ for a planar film) by the
emission of a photon~\cite{1968Mooradian}. The metal fluorescence can largely overlap with the emission spectrum of the
emitter and in some cases may exceed the signal. This in particular is a disadvantage for emitters with a broad
spectrum. For this reason, a chemical reduction process of silver nitrate in solution~\cite{2002Sun} is the preferred
fabrication method for highly crystalline silver nanowires. Since these wires are prepared in solution they can usually
just be placed on the final sample with a spin casting process yielding their position at random. The silver wires
obtained after a washing procedure (and size selection) in a centrifuge are protected by a thin polymer layer. The
polymer prevents Ag from corrosion when exposed to ambient air and can be used as an adhesive layer for particles such
as diamond nano-crystals~\cite{2009Kolesov} or colloidal QDs.

Template stripping, as demonstrated by Nagpal et al.~\cite{2009Nagpal}, with precisely patterned silicon substrates
appears as an alternative and scalable fabrication method for quantum plasmonic circuitry. They patterned a silicon
wafer with a focussed ion beam or lithography. Afterwards the smooth surface was coated with a metal and epoxy. Due to
the bad adhesion between silicon and the metal, the metal epoxy bilayer can be peeled off after deposition yielding a
patterned metallic structure with a surface roughness determined by the substrate. As a proof of the superior surface
quality, they have shown that the propagation length of surface plasmons on planar silver films fabricated with this
method is mainly limited by Ohmic losses, caused by electron scattering with background ions and themselves. The
dependence on the metal deposition parameters in template stripped waveguides was studied by McPeak et
al.~\cite{2015McPeak}, showing that for optimized parameters the guiding properties are comparable to those of highly
crystalline structures.

\subsection{Statistical coupling of single emitters to nanowire waveguides}
The dipole emitter - surface plasmon excitation is an optical near field coupling process which scales as $\sim 1/r^3$
for cylindrically shaped nanowires, where $r$ is the nanowire radius. Silver nanowires synthesized with the
wet-chemical process typically have a radius in the range $10-500 \, nm,$ in which case surface plasmons are most
efficiently excited for $r_d$ between $10 \, nm$ and $50 \, nm$. For very short distances $r_d$ of a few nanometer,
$\Gamma_{non-rad}$ becomes the dominant decay channel independent of the wire diameter~\cite{2006Chang}. In the present
context, $\Gamma_{non-rad}$ refers to the emitter coupling to lossy plasmon modes~\cite{2015Pelton}. In some context
this is called quenching and is not to be mistaken with non-radiative decay processes intrinsic to some emitters.

\begin{figure}[hbt]
    \begin{center}
    \textbf{a) Akimov et al., 2007} \\
    \textbf{b) Kolesov et al., 2009} \\
    \textbf{c) Fedutik et al., 2007} \\
    \end{center}
    \caption[]{Experimental approaches on random single emitter surface plasmon coupling: (a) Single CdSe colloidal QDs
     are placed next to single silver nanowires, separated by a PMMA spacer layer from the nanowire~\cite{2007Akimov}.
    (b) Atomic force microscope image of a silver nanowire with adhered nano-diamonds containing single NV centers~\cite{2009Kolesov}.
    (c) CdSe colloidal QDs placed on a silver nanowire, separated by a $Si O_2$ spacer layer~\cite{2007Fedutik}. \label{fig-emitter_coupling_random}}
\end{figure}
The fist ground breaking result on single-emitter surface plasmon coupling was obtained by Akimov et
al.~\cite{2007Akimov} (Fig.~\ref{fig-emitter_coupling_random} (a)). In their study, the samples were prepared by
spin-casting colloidal CdSe quantum dots (QDs) on glass substrates followed by a layer of poly-methylmethacrylate
(PMMA) with thicknesses in the range $30-100 \, nm$ acting as a separation layer to the silver nanowires, subsequently
deposited using a stamp. Afterwards the samples were covered with another layer of PMMA ensuring a symmetric optical
environment and preventing the silver from corrosion. Single QDs, identified by second-order correlation function
measurements $g^{(2)}(\tau)$ on the emitted photon statistics, were at random found to be coupled to the silver
nanowire surface plasmon mode. For nanowires with a diameter $\sim 102 \pm 24 \, nm$, they observed a mean total decay
rate enhancement of up to 1.7 compared to uncoupled QDs for a separation of $r_d \sim 30 \, nm$. The excitation of the
surface plasmon mode was further verified by the observation of photon re-emission from the distant nanowire end, which
was anti-correlated with direct QD emission as witnessed by a cross-correlation measurement between direct radiative
emission, corresponding to $\Gamma_{rad},$ and the distant nano-wire end, corresponding to excitations of the
propagating plasmonic mode, $\Gamma_{pl}.$ Due to an inhomogeneous distribution of the QD spontaneous emission rate,
only statistical estimates on the decay rate enhancement and the resulting coupling to surface plasmons could be
provided, which is an intrinsic limitation of the spin casting and random assembly approach. In a related experiment,
Fedutik et al.~\cite{2007Fedutik} (Fig.~\ref{fig-emitter_coupling_random} (c)) separated an ensemble of CdSe colloidal
QDs from silver nanowires by coating the wires with a thin $SiO_2$ spacer layer and verified the plasmon excitation by
observing scattered surface plasmons from the wire ends.

The first related experiment with diamond nano-crystals containing single nitrogen-vacancy (NV) centers was reported by
Kolesov et al.~\cite{2009Kolesov} (Fig.~\ref{fig-emitter_coupling_random} (b)). They attached the diamond nanocrystals
in solution to the silver nanowires by making use of the adhesive polymer layer surrounding the silver nanowires. Using
the broad NV center optical emission spectrum of $\approx 100 \, nm$ and the single photon statistics, Kolesov et al.
could verify wave-particle duality of surface plasmon polaritons by observing a strong modulation in the resulting
nanowire spectrum (respectively the plasmonic decay channel $\Gamma_{pl}$) and anti-bunching in the photon number
statistics. Li et al.~\cite{2014Li} deposited silver nanowires on a glass substrate and covered the wires with a $10 \,
nm$ thin layer of $Al_2 O_3$ using an atomic layer deposition technique. For CdSe/ZnS core/shell QDs spin casted on the
sample and thereby deposited next to a nanowire, they reported on surface plasmon excitation with an efficiency up to
$40\%.$

\subsubsection{Controlled coupling of single emitters to nanowires}
The experiments described in the previous section realized emitter - surface plasmon coupling through statistical
assembly. Because of inhomogeneous lifetime broadening of solid state emitters such as colloidal QDs or NV centers in
nano-diamonds, this approach yields a rather significant uncertainty in the estimate of the total decay rate
enhancement and the $\beta-\textrm{factor}.$ To eliminate this uncertainty and to investigate the coupling strength
using only one emitter, it was suggested and demonstrated to follow a different approach utilizing an atomic force
microscope (AFM) as imaging and positioning tool~\cite{2011Huck}. Experimentally, it is convenient to combine the AFM
with the optical setup to allow for the simultaneous acquisition of fluorescence and sample topography images. The
samples were prepared on plasma cleaned fused silica substrates by spin casting diamond nano-crystals (Mikrodiamant MSY
0-0.05) and colloidal silver nanowires from diluted solutions. This sequence yields a uniform distribution of nanowires
and diamond crystals on the substrate. Using a combination of optical characterization techniques (fluorescence
imaging, lifetime, spectrum, and auto-correlation measurement $g^{(2)}(\tau))$ and the sample topography acquired with
the AFM, it is possible to identify individual diamond nanocrystals containing a single NV-center. After switching the
AFM from tapping mode to contact mode operation, pressing the tip with a small force of $\approx 1 \mu N$ on the sample
and manually controlling the tip position, one can isolate single diamond crystals from others and push them towards a
nearby silver nanowire. This procedure works well for crystals with a diameter $>20 \, nm,$ while smaller crystals my
be picked up by the AFM tip~\cite{2011Schell}. For simplifying the moving procedure it is also possible to clean larger
sample areas from particles by scanning the AFM tip in contact mode with a small force across the sample surface. After
approaching the diamond nanocrystal containing the single NV center and a silver nanowire, another lifetime measurement
is taken. The total decay rate enhancement can then be determined by comparing the lifetime before and after coupling
to the nanowire. A correlation function measurement with a value of $g^{(2)}(\tau = 0) < 0.5$ taken on the coupled
system ensures that the signal originates from the NV center and not from unwanted background fluorescence. With
typical nanowire diameters in the range $30-100 \, nm$ and diamond crystals with a mean and maximum size of $35 \, nm$
and $50 \, nm,$ respectively, a total decay rate enhancement in the range $2-4$ was commonly observed~\cite{2011Huck}.
Surface plasmon excitation is further verified, similar to the work by Akimov at al.~\cite{2007Akimov}, by the
observation of surface plasmon scattering to the radiation field at the distant nanowire ends.

In continuation of these achievements, it would be natural to further improve the coupling to surface plasmon modes
using smaller diamond crystals (reducing $r_d$) and thinner nanowires (improving the mode confinement). Smaller diamond
crystals containing single NV centers~\cite{2009Tisler} are attractive in this context because they also reduce the
uncertainty on the coupling distance $r_d$ partially determined by the size of the diamond crystal.

In a more general context other propagating plasmonic modes such as channel plasmon polaritons
(CPPs)~\cite{2004Pile,2005Pile,2006Bozhevolnyi} occurring for instance in V-grove sculptured metallic films or hybrid
gap modes localized between parallel nanowires~\cite{2009Manjavacas} were explored experimentally. Compared to single
nanowires, CPP and gap modes offer the advantage that their effective mode area is largely reduced, the plasmon field
maximum is accessible by an emitter and the fields propagate over relatively long distances.

\subsubsection{Controlled coupling to a plasmon gap mode}
Due to the finite size of the diamond crystals and the limitation in fabricating thinner single silver nanowires, the
possibility to excite the highly confined surface plasmon gap modes occurring between two parallel silver nanowires was
explored~\cite{2013Kumar}. In parallel configuration of two nanowires with a small gap, the single modes cease to exit
and two hybrid modes form. These hybrid modes can be understood as an in-phase and out of phase superposition of the
single modes and are referred to as symmetric (+,+) or anti-symmetric (+,-) modes, depending on the transvers
charge/phase distribution. The improved field confinement and the increased coupling rates of the gap plasmon mode
compared to a single nanowire mode are summarized in Fig.~\ref{fig-gap-modes-benifit}.
\begin{figure}[hbt]
    \begin{center}
    \textbf{a) and b) Kumar et al., 2013}
    \end{center}
    \caption[]{Pointing vector of the guided surface plasmon mode on (a) a single nanowire and (b) the anti-symmetric
    gap mode with a separation of $9\, nm$ between the wires. (c) Comparison of the normalized decay rate into the
    plasmon mode $\Gamma_{pl}/\Gamma_0$ between the single and the dual nanowire configuration, parameterized as a
    function of linear propagation loss $\kappa$. All graphs are taken from~\cite{2013Kumar}.
    \label{fig-gap-modes-benifit}}
\end{figure}
It is the large mode confinement in the gap region and the smaller propagation losses $\kappa$ compared to a single
silver nanowire making this structure attractive for emitter plasmon coupling. The anti-symmetric $(+,-)$ mode
facilitates a large plasmonic decay rate $\Gamma_{pl}$ when the emitter is placed at the mode field maximum in the gap
region, which is about one order of magnitude larger compared to maximum achievable decay rates to single nanowires
with similar linear propagation losses $\kappa$ (Fig.~\ref{fig-gap-modes-benifit} (c)).

The experiment reported by Kumar et al.~\cite{2013Kumar} directly demonstrates the dual nanowire advantage by selecting
a NV center with a relatively long intrinsic lifetime of $45.2 \, ns,$ which reduced to $11.9 \, ns$ after coupling to
a single nanowire and further down to $5.4 \, ns$ when a second nanowire was placed nearby
(Fig.~\ref{fig-results-gap-modes}). The diamond nano-crystal was measured with a height of $27 \, nm$ which is
significantly smaller than the nanowire radius of $55 \, nm.$ Hence, the NV center was not located at the maximum
electric field, $max\{\mathbf{E}(\mathbf{r}) \},$ in the gap region. An optical image of the final structures
highlights that the integrated photon count rate from the nanowire ends exceeded the collected radiative emission from
the NV center.
\begin{figure}[hbt]
    \begin{center}
    \textbf{a), b) and c) Kumar et al., 2013}
    \end{center}
    \caption[]{(a) Lifetime measurements of one NV center when located on the glass substrate (black), after coupling
    to a single nanowire (red) and after locating in the gap between two nanowires (blue). (b) Fluorescence image of the dual wire
    structure. \label{fig-results-gap-modes}}
\end{figure}

\subsubsection{NV center coupling to a channel plasmons in a V-grove}
Recently, Berm\'{u}dez-Ure\~{n}a et al. coupled single NV centers to the channel plasmon mode of a V-groove waveguide
~\cite{2015Urena} (Fig.~\ref{fig-V-groove}). V-groves with a width $\sim 315 \, nm$ and a hight $\sim 510 \, nm$ were
milled with a focused ion beam into a gold film of $1.2 \mu m$ total thickness~\cite{2006Bozhevolnyi}. On their ends
the V-groves were terminated with tapered nano-mirrors~\cite{2011Radko}, as shown in Fig.~\ref{fig-V-groove} (b). After
depositing an array of nano-diamonds with a controlled electron beam lithography method nearby the V-groove, a suitable
NV center with a long lifetime ($\sim 26 \, ns$) and high count rate was selected and placed in the groove by the aid
of an AFM tip. The channel plasmon propagation length was measured to be~$4.65 \pm 0.48 \mu m$ using the the NV center
coupling to the plasmon mode, matching the theoretically expected value of $4.56 \mu m$ obtained by averaging over the
broad NV center spectrum. Comparing the NV center lifetimes before and after coupling yields a total decay rate
enhancement of $2.3$ and together with an estimates of the propagation losses and coupling efficiency to the radiation
field at the groove end, a $\beta-\textrm{factor}$ of $0.42 \pm 0.03$ was estimated, which is in good agreement with
simulations indicating a value of $\sim 0.56$ for this structure.

\begin{figure}[hbt]
    \begin{center}
    \textbf{a), b), and c), Urena et al., 2015}
    \end{center}
    \caption[]{(a) Illustration of the approach for NV center channel plasmon coupling: A green laser excites the NV center
    which decays by exciting channel plasmons along the V-groove and subsequently scatter to the far field at the ends. (b)
    Scanning electron microscope image of a V-groove fabricated in gold, illustrating the groove profile and a mirrors.
    (c) Total electric field profile of the supported channel plasmon mode. The field polarization is indicated by arrow.
    Images taken from~\cite{2015Urena}.  \label{fig-V-groove}}
\end{figure}

\subsubsection{Alternative methods for deterministic coupling}
As the AFM assisted assembly is limited to relatively large crystals, is rather time consuming and might even fail in
some cases, alternative methods are being pursued aiming at deterministic position control of an emitter to metallic
waveguides. Gruber et al.~\cite{2012Gruber} applied a two-step electron beam lithography process to first fabricate
silver nanowires and second deposit a small number of colloidal QDs at the nanowire ends by spin-casting.
Alternatively, Pfaff et al.~\cite{2009Pfaff} first determined the position of NV centers on a $SiO_2$ substrate with
respect to alignment marks and afterwards fabricated Ag and Al nanowires on top with an electron-beam lithography
process. A microfluidic device was used by Ropp et al.~\cite{2013Ropp} for positioning and moving QDs around single
silver nanowires, enabling them to map out spontaneous emission modifications with a corresponding $12 \, nm$ spatial
accuracy of the QD.

\subsection{Plasmonic resonators}
\begin{figure}[hbt]
    \begin{center}
    \textbf{a) Leon et al., 2012} \\
    \textbf{b) Kress et al., 2015} \\
    \end{center}
    \caption[]{Plasmon resonators made (a) on a silver colloidal nanowire by defining DBR mirrors on the surrounding
    PMMA medium~\cite{2012Leon} and (b) by adding block reflectors confining a plasmon mode propagating along the
    wedge~\cite{2015Kress}. The scale bars in (a) and (b) both correspond to $1 \, \mu m.$
    \label{fig-plasmon-resonator}}
\end{figure}
A nanoscale plasmonic resonator based on synthetic silver nanowires was proposed and demonstrated by de Leon et
al.~\cite{2012Leon}. The silver nanowires were embedded in PMMA, and a cavity resonance was achieved by defining
distributed Bragg reflectors (DBRs) at the nanowire ends with electron beam lithography (see
Fig.~\ref{fig-plasmon-resonator} (a)). The achieved DBR plasmon reflection was in the range $90-95\%,$ resulting in a
measured Q-value of $58$ (highest value they report on is $94$) at a vacuum wavelength of $638\,nm$ for a nanowire with
a diameter of $\sim 100 \, nm$, close to the theoretical expectation of $Q \sim 100$. Together with a small effective
mode of $V_{eff} \sim 0.04 (\lambda_0/n)^3$ they expect a Purcell factor up to $200$ and measure $F>75$ with CdSe
colloidal QDs by comparing the lifetimes of coupled and uncoupled QDs. For NV centers in diamond nanocrystals a Purcell
factor of $\sim 35$ is reported, exceeding the value of dielectric cavities~\cite{2010Englund, 2011vanderSar}.

Wedge waveguides and resonators, made with template stripping for achieving long propagation~\cite{2009Nagpal}, have
recently been investigated by Kress et al.~\cite{2015Kress} for enhanced light matter interaction at the single emitter
level. Normalized to Ohmic propagation losses, wedge waveguides show the highest field
confinement~\cite{2008OultonConf,2008Moreno} compared to other structures such as channels~\cite{2006Bozhevolnyi},
nanowires~\cite{1997Takahara} and hybrid plasmonic waveguides~\cite{2008Oulton}. The resonators were made by adding
block reflectors across the wedge having a reflectivity of $\sim 93\%$ (see Fig.~\ref{fig-plasmon-resonator} (b)). With
the long propagation length of $19 \, \mu m$ (at a vacuum wavelength of $630 \, nm$), they measured a Q-value $\sim
191$ with a $10 \mu m$ long resonator. Using a modified electrostatic printing technique~\cite{2012Galliker} they
deposited core/shell/shell CdSe/CdS/ZnS colloidal QDs on the wedge~\cite{2014Kress} and the lifetime reduced by a
factor of $22.6$ compared to emitters dispersed in liquid.

\section{Quantum applications of strongly confined propagating plasmons}
Systems of propagating photons in plasmonic structures with single emitters open the potential for efficient single
photon generation, single photon absorption and strong photon-photon interaction mediated by a strong non-linearity. As
a result of the potentially large Purcell effect of a single emitter coupled to a plasmonic wire, photons spontaneously
emitted from the dipole emitter will be harvested by the propagating plasmon mode of the wire and thus directed into a
single well-defined spatial mode. By coupling this plasmonic mode to the mode of a dielectric waveguide, it has been
shown theoretically that single photons can be generated with high efficiency and high speed~\cite{2006Chang}.

An even more appealing application of the plasmon-emitter system is the promise of a giant non-linearity that enables a
strong interaction between individual plasmons~\cite{2014Chang}. This may impact the fields of nonlinear optics and
quantum optics. For example, the strong interaction may lead to the realization of an optical switch at the single
photon level~\cite{2007ChangTrans}. The basic idea is to make use of the saturation nonlinearity of a two-level
emitter: The emitter will absorb and thus scatter off a single photon while it will be invisible to the next photon (as
it has been excited by the first one). This makes up the single photon switch. A similar strategy using a three-level
system in replacement of the two-level system, may lead to the realization of a single photon transistor which might
have important usage in quantum information processing and in quantum networks~\cite{2008Kimble}.

On the more exotic side, it has been predicted that the combination of strong non-linear interaction and directional
coupling between a larger number of emitters may lead to multi-partite entanglement as well as to new quantum phase
transitions of light~\cite{2006Hartman,2006Greentre,2007Angelakis} or photon
crystallization~\cite{2008Chang,2013Otterbach}. Originally, such proposals have envisioned the use of cavity quantum
electrodynamics in high-finesse cavities~\cite{1995Turchette,2005Birnbaum}, but more recently the ideas have been
formulated also in the context of propagating modes coupled to emitters in one-dimensional waveguides~\cite{2008Chang}.
In addition to using plasmonic waveguides~\cite{2011Tudela,2013Tudela,2013Hummer}, there has been proposals on using
atomic clouds coupled to guided modes of photonic waveguides such as tapered fibers~\cite{2010Vetsch,2012Goban},
photonic crystal fibers~\cite{2009Bajcsy} or photonic crystal waveguides~\cite{2014Goban}. The ideas have been also
translated into systems based on superconducting qubits coupled to microwave waveguides~\cite{2013vanLoo,2013Hoi}.

\section{Outlook}
Most of the experiments on a plasmonic platform exhibiting true quantum properties have been based on simple circuitry
such as a single wire, two coupled wires, V-groves, or wedges. However, to make the platform attractive for quantum
applications, it is important to be able to make more sophisticated plasmonic circuits that include the plasmon
generation process, the linear and non-linear interaction and potentially also the detection process on-chip. As we
have seen above, relatively efficient generation of single photons have been demonstrated, beam-splitting and
interference has been shown, and on-chip detection has been realized~\cite{2009Falk}. However, the complete
demonstration of all three stages on a single chip has not been realized yet.

An important first test towards quantum physics in a plasmonic system is the observation of Hong-Ou-Mandel interference
between two single photons generated on-chip from two independent single photon emitters and with two on-chip
detectors. Initial steps towards such a landmarking experiment have been performed where the plasmonic interference was
observed between two externally generated photons that were coupled onto the chip for interference and outcoupled again
for detection~\cite{2014Fakonas,2014Martino}. Another experiment with an external source but with on-chip detection has
shown indication of Hong-Ou-Mandel interference~\cite{2013Heeres}. However, the plasmonic circuitry for all these
demonstrations are not optimized for coupling to single emitters, and thus for a complete demonstration where all
components are on-chip will require the design and development of a new system. A first critical step is to demonstrate
plasmon-plasmon interference with two on-chip emitters but using external detectors. This can for example be carried
out using two silicon-vacancy (SiV) centers in diamond which have shown to exhibit quantum photon interference without
the need for special filtering or frequency control~\cite{2014Sipahigil}. After such a milestone demonstration, the
next step would be full integration with emitters, manipulation and detection paving the way for more complex nanoscale
quantum plasmonic devices.

The construction of such a plasmonic quantum chip is fraught with technical challenges associated with fabrication and
integration. However, one of the biggest challenges that we are faced with when trying to scale the circuits is loss.
The loss-problem simply has to be solved to make it a viable platform for scalable quantum information processing since
propagation loss limits the capabilities in carrying out fault-tolerant quantum information processing. In classical
information processing, losses can be overcome by the insertion of on-chip amplifiers that compensate for the
losses~\cite{1986Maers,1987Maers}, and demonstrated by Noginov et al. for propagating surface plasmon
modes~\cite{2008Noginov}. Such amplification processes cannot be used in quantum information processing as it
inevitably will add noise to the plasmonic quantum states~\cite{1961Louisell,1962Haus,1982Caves}, thereby destroyingthe
quantum information. As an alternative, one needs to devise alternative designs that minimize the losses. Above we
discussed a couple of systems based on gap plasmons in which the loss rate could be reduced while the high coupling
strength remained at the same level. To reduce the loss rate even further, hybrid approaches have been introduced. Such
approaches are combined systems of plasmonic and photonic waveguides where one tries to make use of the low loss
properties of the photonic waveguides and the strong confinement properties of the metallic
waveguides~\cite{2008Oulton}. Yet another alternative is to replace the noble metal structures with other materials,
for example ceramic compounds~\cite{2014Boltasseva}, which relative to silver and gold have shown reduced losses in the
near and mid infrared regimes where several semiconductor systems are active. Further development of these and similar
ideas will be critical to the success of building up larger quantum plasmonic circuits for quantum information
processing. However, quantum plasmonic circuitry is still in its infancy and it can be expected that new discoveries
might be uncovered and lead to new key turning points in the engineering of low-loss plasmonic circuits.

\end{document}